\documentclass[%
 reprint,
superscriptaddress,
 amsmath,amssymb, aps,
]{revtex4-1}

\usepackage{graphicx}
\usepackage{dcolumn}
\usepackage{bm}

\begin{document}

\preprint{APS/123-QED}

\title{Transition from quasi-unidirectional to unidirectional guided resonances in leaky-mode photonic lattices}%

\author{Sun-Goo Lee}
\email{sungooleee@gmail.com}
\affiliation{Department of Data Information and Physics, Kongju National University, Gongju, 32588, Republic of Korea}%
\affiliation{Institute of Application and Fusion for Light, Kongju National University, Cheonan, 31080, Republic of Korea}%

\author{Seong-Han Kim}
\affiliation{Advanced Photonics Research Institute, Gwangju Institute of Science and Technology, Gwangju 61005, Republic of Korea}%

\author{Wook-Jae Lee}
\affiliation{Department of Data Information and Physics, Kongju National University, Gongju, 32588, Republic of Korea}%
\affiliation{Institute of Application and Fusion for Light, Kongju National University, Cheonan, 31080, Republic of Korea}%

\date{\today}

\begin{abstract}
Unidirectional light emission from planar photonic structures is highly advantageous for a wide range of optoelectronic applications. Recently, it has been demonstrated that unidirectional guided resonances (UGRs) can be realized by utilizing topological polarization singularities in momentum space. However, the practical application of these topological unidirectional emitters has been limited due to their intricate geometric configurations, requiring special efforts with high-cost fabrication processes. In this study, we show that unidirectional light emission can be achieved in conventional one-dimensional zero-contrast gratings (ZCGs), which can be easily fabricated using current nanofabrication technologies. In ZCGs, the interband coupling between even-like and odd-like waveguide modes leads to the formation of quasi-UGRs, characterized by significantly higher decay rates in either the upward or downward direction compared to the opposite direction. We demonstrate that these quasi-UGRs evolve into genuine UGRs with an gradual increase in grating thickness. Moreover, the emission direction of UGRs can be selectively steered either upward or downward by adjusting the lattice parameters. In addition to quasi-UGRs and UGRs, our study also reveals additional topological phenomena in ZCGs, including exceptional points and quasi-BICs.
\end{abstract}
\maketitle


\section{Introduction}
Unidirectional light emission from planar photonic structures, with no emission in the opposite direction, holds crucial importance for a wide range of practical applications as well as in fundamental scientific research \cite{HZhou2016,STHa2018}. Conventionally, to prevent light emission in undesired directions, perfect reflectors such as metallic mirrors or photonic band gap materials are employed \cite{Roncone1993,Taillaert2004,YOta2015}. However, the integration of additional reflectors into optical devices leads to increased bulkiness and adds complexity to the fabrication process. Consequently, this renders conventional unidirectional emitters less suitable for compact and planar on-chip photonic applications. Unidirectional Guided Resonances (UGRs) in planar photonic lattices refer to specific resonant optical eigenstates that emit light exclusively in one direction. This distinctive directional radiation stems from the inherent topological characteristics of these resonances. \cite{XYin2020,ZZhang2021,ZZhang2022}. Unlike traditional methods, these topologically enabled UGRs obviate the need for additional reflectors, presenting a streamlined approach for unidirectional light emission. This advancement paves the way for their application in a variety of optoelectronic devices, including surface-emitting lasers \cite{MCYHuang2007,HMatsubara2008,Kodigala2017}, optical antennas \cite{MRaval2017,SKhajavi2021,XZhang2022}, and highly efficient grating couplers \cite{HWang2023,MDai2015,AMichaels2018}.

UGRs have been observed in various periodic photonic structures, including one-dimensional (1D) gratings and two-dimensional (2D) photonic crystal slabs. This phenomenon is closely linked to topological polarization singularities in momentum space, such as optical bound states in the continuum (BICs) \cite{Marinica2008,Plotnik2011,Hsu2016,Bulgakov2017,Koshelev2018,Koshelev2019,SGLee2019-1,WLiu2019,TYoda2020,SGLee2021-1} and circularly-polarized states ($C$ points) \cite{Schoonover2006,MBurresi2009,Fosel2017}. BICs, acting as vortex centers with quantized integer topological charges, exhibit infinite radiative $Q$ factors due to their non-radiative nature \cite{BZhen2014,Doeleman2018,JJin2019,MKang2022-1,SGLee2023}. In contrast, UGRs, also acting as vortex centers, carry topological charges but with finite $Q$ factors, enabling their unidirectional radiation. UGRs typically arise at specific $k$-points where multiple $C$ points, ecah carrying half-integer topological charges, converge \cite{XYin2020,YZeng2021}. Traditionally, the generation of $C$ points in photonic crystal slabs necessitates disrupting the in-plane $C_2$ symmetry. Therefore, to achieve UGRs in photonic crystal slabs, it is essential to eliminate both the in-plane $C_2$ symmetry and the up-down mirror symmetry \cite{XYin2020}. This leads to complex unit cell geometries and expensive fabrication processes. However, a recent study has revealed that topological $C$ points can be generated through interband coupling between different waveguide modes, even without removing the in-plane $C_2$ symmetry \cite{XYin2023}. The UGRs induced by interband coupling offer practical advantages. The less stringent symmetry requirements allow for a more straightforward structural design of unit cells, thereby simplifying the fabrication of topologically enabled UGRs.

\begin{figure*}[]
\centering\includegraphics[width=16 cm]{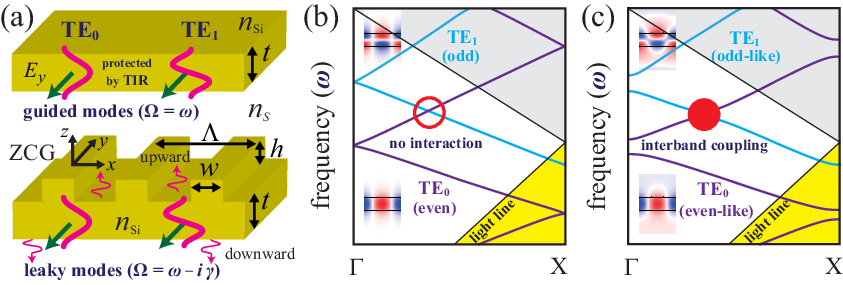}
\caption {\label{fig1} (a) Schematic representation of a homogeneous dielectric waveguide and a ZCG with broken up-down mirror symmetry. While leaky modes in the ZCG are characterized by complex eigenfrequencies due to continuous electromagnetic energy loss, the guided modes in the homogeneous waveguide exhibit real eigenfrequencies, indicating energy conservation within the system. (b) Dispersion curves for $\mathrm{TE}_{0}$ and $\mathrm{TE}_{1}$ modes in a homogeneous dielectric waveguide. The insets with blue and red colors depict the spatial distribution of electric fields for the even and odd modes. (c) A conceptual illustration of photonic band structures for a ZCG. Interband coupling between an even-like $\mathrm{TE}_{0}$ mode and an odd-like $\mathrm{TE}_{1}$ mode gives rise to the interesting topological physical phenomena in the region marked with a red circle.}
\end{figure*}

In this paper, we demonstrate that conventional 1D zero-contrast gratings (ZCGs) can function effectively as unidirectional light emitters by harnessing topologically enabled UGRs. ZCGs, which exhibit broken up-down mirror symmetry while retaining in-plane $C_2$ symmetry, are compatible with existing nanofabrication technologies \cite{Magnusson2014,MNiraula2015}. In ZCGs, interband coupling between even-like and odd-like waveguide modes leads to the emergence of quasi-UGRs. These quasi-UGRs are characterized by significantly higher decay rates in either the upward or downward direction compared to the opposite direction. Notably, the quasi-UGRs exhibit robustness against variations in lattice parameters and can be readily identified in 1D momentum space. We reveal that quasi-UGRs gradually evolve into genuine UGRs as the grating thickness increases. Moreover, the emission direction of these UGRs raised by interband coupling can be selectively controlled to be either upward or downward by tuning the lattice parameters. While previous studies have focused on tracking $C$ points in 2D momentum space to create topologically enabled UGRs \cite{XYin2020,YZeng2021}, our approach in 1D momentum space significantly reduces computational demands. We further show that our method, centered on the evolution of quasi-UGRs in 1D momentum space, is applicable for observing topologically enabled UGRs in a variety of photonic lattices, including those lacking in-plane $C_2$ symmetry. Besides quasi-UGRs and UGRs, we also observe other topological phenomena in ZCGs, such as exceptional points and quasi-BICs.
 
Figure~\ref{fig1}(a) presents a illustration of a planar homogeneous waveguide and a conventional 1D ZCG for the study of interband-coupling-induced UGRs. Both the waveguide and the ZCG are situated in a medium with a refractive index of $n_s=1.46$ and are composed of silicon ($\mathrm{Si}$) with a refractive index $n_\mathrm{Si}=3.48$ at a wavelength $\lambda =1.55~\mu \mathrm{m}$. The ZCG is constituted by a uniform waveguide with a thickness of $t$ and a grating composed of rectangular-shaped $\mathrm{Si}$ rods with a period $\Lambda$. The width and height of $\mathrm{Si}$ rods are denoted as $w$ and $h$, respectively. In the homogeneous $\mathrm{Si}$ waveguide, guided modes are perfectly protected by total internal reflection (TIR) and propagate along the waveguide without out-of-plane radiation \cite{Agrawal2004}. Figure~\ref{fig1}(b) depicts the dispersion relations of the two lowest transverse electric modes $\mathrm{TE}_{0}$ and $\mathrm{TE}_{1}$. For convenience, the dispersion relations are plotted within the reduced Brillouin zone corresponding to a 1D grating with a period of $\Lambda$. Since the $\mathrm{TE}_{0}$ and $\mathrm{TE}_{1}$ modes have perfectly even and odd transverse electric field profiles, respectively, and are orthogonal to each other, they do not interact with each other, allowing their dispersion curves to cross. The ZCG with a height of $h=0$ (or width of $w=0$) is equivalent to a uniform waveguide with a thickness of $t$. As $h$ (or $w$) increases slightly from zero, with $w\neq 0$ (or $h\neq 0$), the $\mathrm{TE}_{0}$ and $\mathrm{TE}_{1}$ modes in the ZCG begin to exhibit even-like and odd-like field distributions, respectively, and start to interact with each other. In the uniform waveguide, guided modes possess real eigenfrequencies, where $\Omega = \omega$. In contrast, leaky modes in the ZCG exhibit complex eigenfrequencies, given by $\Omega = \omega - i \gamma$. The radiative energy loss of the leaky modes is quantified by the quality factor, defined as $Q = \omega / 2\gamma$ \cite{Joannopoulos2008}. 

When two nonorthogonal waveguide modes, $\mathrm{TE}_{0}$ and $\mathrm{TE}_{1}$, with eigenfrequencies $\Omega_0$ and $\Omega_1$, interact with each other, the Hamiltonian for the coupled modes, according to coupled-mode theory \cite{WSuh2004,SGLee2020-1}, is given by
\begin{equation}\label{Hamiltonian}
\mathcal{H} = \left [ \begin{matrix} \Omega_{0} & \alpha  \\ \alpha  & \Omega_{1} \\ \end{matrix} \right ] -i \left [\begin{matrix} 0 & \beta \\ \beta & 0 \\ \end{matrix} \right], \\
\end{equation} 	 	
where $\alpha$ and $\beta$ denote the near-field and far-field coupling between the waveguide modes, respectively. The two eigenstates of $\mathcal{H}$ can be written as 
\begin{equation} \label{eigenstates}
\Psi^{+,-}(k_x) = c^{+,-}_0 \psi_0(k_x) + c^{+,-}_1 \psi_1(k_x),
\end{equation} 	 	
where $\psi_{0,1}$ represents the eigenstate of the $\mathrm{TE}_{0,1}$ mode without coupling, and $[c_{0}, c_{1}]^{\rm{T}}_{+,-}$ denotes the eignenvector of $\mathcal{H}$, with eigenvalues given by
\begin{equation}\label{eigenvalue}
\Omega^{+,-} = \frac{\Omega_0+\Omega_1}{2}\pm\sqrt{\frac{\left ( \Omega_0-\Omega_1\right )^2}{4}  + (\alpha - i \beta)^2}.
\end{equation} 	
Due to interband coupling, in the region marked with a red circle in Fig.~\ref{fig1}(c), eigenvalues can be substantially altered from their original values, leading to the observation of topological phenomena such as exceptional points, quasi-BICs, quasi-UGRs, and UGRs.

In this study, we investigate the topological physical phenomena that arise from the interband coupling between the even-like $\mathrm{TE}_{0}$ mode and the odd-like $\mathrm{TE}_{1}$ mode. We identify exceptional points and quasi-BICs by examining the dispersion relations and the associated radiative $Q$ factors. To quantify the directionality of radiation in the coupled eigenstates, we employ the radiation ratio, defined as $\eta = \gamma_u / \gamma_d$. Here, $\gamma_u$ and $\gamma_d$ denote the decay rates in the upward ($+z$) and downward ($-z$) directions, respectively. Within this framework, eigenstates exhibiting $\eta \geq 80~\rm{dB}$ are defined as UUGRs, primarily emitting light upwards, while those with $\eta \leq -80~\rm{dB}$ are identified as DUGRs, mainly radiating downwards. We introduce the concepts of quasi-UUGRs and quasi-DUGRs to further elucidate the origins of UUGRs and DUGRs. The values of $\eta$ for quasi-UUGRs are characterized by $\eta < 80~\rm{dB}$, while those for quasi-DUGRs are characterized by $\eta > -80~\rm{dB}$. Notably, in the $\eta$ curves plotted as a function of $k_x$, we observe apparent local maxima at points corresponding to quasi-UUGRs and local minima at those corresponding to quasi-DUGRs. To investigate topological physical phenomena in ZCGs, we perform rigorous finite-element method (FEM) simulations using the commercial software COMSOL Multiphysics. We analyze the dispersion relations, radiative $Q$ factors, and radiation ratio $\eta$ by varying the lattice parameters.

\section{Results}
\begin{figure*}[]
\centering\includegraphics[width=16.5 cm]{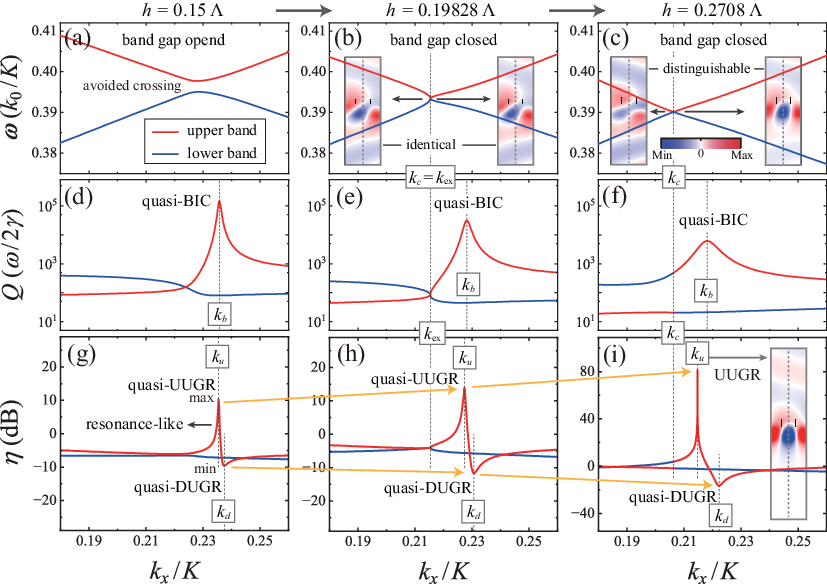}
\caption {\label{fig2} Exceptional points, quasi-BICs, quasi-UUGRs, quasi-DUGRs, and UUGRs resulting from the interband coupling between the even-like $\mathrm{TE}_{0}$ mode and odd-like $\mathrm{TE}_{1}$ mode in ZCGs. The figures showcase the evolution of dispersion curves (a)--(c), radiative $Q$ factors (d)--(f), and the asymmetric radiation ratio $\eta$ (i)--(k) under variation of $h$. Quasi-BICs, quasi-UUGRs, and quasi-DUGRs are clearly identifiable in ZCGs. Notably, quasi-UUGRs evolve into UUGRs with the gradual increase in $h$. Insets with blue and red colors in figures (b) and (c) display the spatial electric field ($E_y$) distribution for the degenerate eigenstates at $k_x/K=k_c$. The inset in figure (i) illustrates the spatial $E_y$ distribution for the UUGR at $k_x/K = k_u=0.214731$. The structural parameters $t=0.5~\Lambda$ and $w=0.4~\Lambda$ were fixed in the FEM simulations. }
\end{figure*}

Figure~\ref{fig2} illustrates the evolution of dispersion curves, radiative $Q$ factors, and radiation ratio $\eta$ as the grating thickness $h$ varies, with fixed values of $t=0.5~\Lambda$ and $w=0.4~\Lambda$. As schematically depicted in Fig.\ref{fig1}(b), in the absence of interband coupling ($h=0$), the dispersion curves for $\mathrm{TE}_{0}$ and $\mathrm{TE}_{1}$ modes intersect as nearly straight lines. However, as the value of $h$ increases from zero, indicating the presence of interband coupling ($h>0$), a photonic band gap opens at the avoided crossing, and the gap size initially increases. However, with further increases in $h$, the gap size diminishes and eventually closes. At $k_x/K=k_c$, the real parts of the eigenfrequencies for the eigenstates in the upper and lower bands become identical. Although the specific value of $k_c$ shifts with further increases in $h$, the bands remain closed. A photonic band gap (at $h=1.5~\Lambda$) and closed band states (at $h=0.19828~\Lambda$ and $0.2708~\Lambda$) can be verified in the dispersion curves presented in Figs.~\ref{fig2}(a)--\ref{fig2}(c).

We now show that ZCGs can exhibit various topological physical phenomena, including exceptional points, quasi-BICs, quasi-UGRs, and UGRs. The radiative $Q$ factors, as depicted in Figs.~\ref{fig2}(d)--\ref{fig2}(f), indicate that ZCGs do not support off-$\Gamma$ BICs with an infinite $Q$ factor. Due to the absence of up-down mirror symmetry, only quasi-BICs with finite $Q$ factors being observed at wave vectors $k_x/K=k_b\neq 0$. Previous work has shown that BICs induced by interband coupling are robust against variations in lattice parameters \cite{SGLee2020-1}. Similarly, our simulation results, including those presented in Figs.~\ref{fig2}(d)--\ref{fig2}(f), demonstrate that quasi-BICs induced by interband coupling also exhibit considerable stability. However, we observe a gradual decrease in the $Q$ factors of quasi-BICs as the grating thickness $h$ increases, indicating that the perturbations become more pronounced with an increase in $h$. Notably, at $h=0.19828~\Lambda$, where the perturbation reaches an appropriate strength, an exceptional point is identified in the momentum space. At $k_x/K=k_{\rm{ex}}=0.215385$, both the real and imaginary parts of the eigenfrequencies coalesce, as shown in Figs.~\ref{fig2}(b) and \ref{fig2}(e), thereby forming an exceptional point. The insets in Fig.~\ref{fig2}(b) demonstrate that the eigenstates at this exceptional point possess identical spatial electric field distributions. In contrast, within the closed band states away from the exceptional point, the electric field distributions of the eigenstates are distinct at $k_x/K=k_c$, as shown in the insets of Fig.~\ref{fig2}(c). Additionally, due to the in-plane mirror symmetry along the $x$-direction in the ZCGs under study, a corresponding exceptional point is also observed at $k_x/K=-k_{\rm{ex}}$.

 \begin{figure*}[]
\centering\includegraphics[width=16.5 cm]{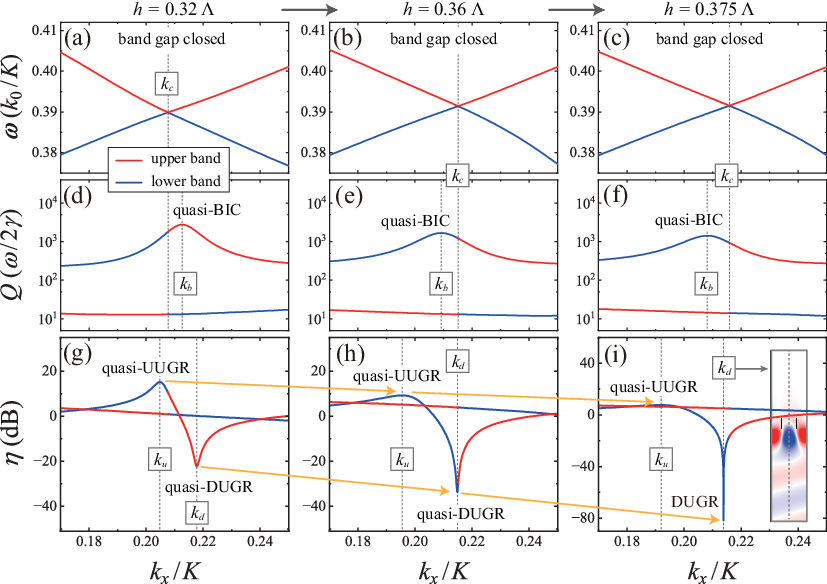}
\caption {\label{fig3} The formation of a genuine DUGR due to the interband coupling between the even-like $\mathrm{TE}_{0}$ mode and odd-like $\mathrm{TE}_{1}$ mode in ZCGs. The evolution of dispersion curves (a)--(c), radiative $Q$ factors (d)--(f), and the asymmetric radiation ratio $\eta$ (i)--(k) as the grating thickness $h$ is varied. Notably, quasi-BICs, quasi-UUGRs, and quasi-DUGRs transition from the red-colored upper band to the blue-colored lower band with incremental changes in $h$. Quasi-DUGRs eventually evolve into DUGRs. The inset in figure (i) illustrates the spatial $E_y$ distribution for the DUGR at $k_x/K = k_d=0.213862$.}
\end{figure*}

By analyzing the radiation ratio $\eta$ as a function of the wave vector $k_x$, we can confirm the presence of interband-coupling-induced quasi-UGRs and UGRs. As illustrated in Fig.~\ref{fig2}(g), at a grating thickness of $h=0.15~\Lambda$, there is a notable difference between upward and downward radiation. Notably, the red-colored $\eta$ curve for the upper band presents an asymmetric resonance-like profile, sharply transitioning from a maximum of $10.2~\rm{dB}$ to a minimum of $-9.6~\rm{dB}$ as $k_x /K $ changes from 0.23549 to 0.23742. We define the eigenstates corresponding to the highest and lowest local $\eta$ values as quasi-UUGRs and quasi-DUGRs, respectively. As depicted in Fig.~\ref{fig2}(g), the quasi-UUGR and quasi-DUGR emit approximately ten times more electromagnetic energy in the upward and downward direction, respectively. At a grating thickness of $h=0.15~\Lambda$, the interband coupling is relatively weak, and no UGR with $|\eta|\geq 80~\rm{dB}$ is observed within the simulated wave vector range $k_x /K \in [0.18, 0.26]$. However, as $h$ increases beyond $0.15~\Lambda$, the strength of interband coupling intensifies, and the maximum (minimum) $\eta$ value for quasi-UUGRs (quasi-DUGRs) increases (decreases), reaching $13.8~\rm{dB}$ ($-11.7~\rm{dB}$) when $h=0.19828~\Lambda$, as shown in Fig.~\ref{fig2}(h). Figure~\ref{fig2}(h) also indicates that the two degenerate eigenstates at the exceptional point $k_x/K=k_{\rm{ex}}$ have identical $\eta$ values. By increasing $h$ further to $0.2708~\Lambda$, as shown in Fig.~\ref{fig2}(i), we observe a UUGR with an $\eta$ of  $81.4~\rm{dB}$ at $k_x/K = k_u=0.214731$. The simulated spatial electric field distributions, depicted in the inset of Fig.~\ref{fig2}(i), confirm that this UUGR predominantly radiates in the upward direction. Moreover, another UUGR is observed at $k_x /K =-0.214731$, consistent with the in-plane $C_2$ symmetry of the ZCGs.

Our simulation results, as presented in Figs.~\ref{fig2}(g)--\ref{fig2}(i), show that quasi-UUGRs and quasi-DURGs are robust against variations in the lattice parameter $h$. Notably, the quasi-UUGRs evolve into UUGRs with a gradual increase in $h$. Figure~\ref{fig2} also confirms that quasi-BICs, quasi-UUGRs, quasi-DUGRs, and UUGRs are all situated in the upper red-colored band. We will now discuss the band transitions of quasi-BICs, quasi-UUGRs, and quasi-DUGRs, as well as the emergence of DUGRs. Figure~\ref{fig3} details the evolution of the band structures, radiative $Q$ factors, and radiation ratio $\eta$ curves as $h$ is increased beyond $0.2708~\Lambda$. With the interband coupling between the $\mathrm{TE}_{0}$ and $\mathrm{TE}_{1}$ modes, as shown in Figs.~\ref{fig3}(a)--\ref{fig3}(c), we observe closed band states at $h$ values of $0.32\Lambda$, $0.36~\Lambda$, and $0.375~\Lambda$. Figures~\ref{fig3}(d)--\ref{fig3}(f) illustrate a decrease in the $Q$ factors of quasi-BICs with increasing $h$, which is attributed to stronger perturbations. Notably, we observe a transition of quasi-BICs from the upper red-colored band to the lower blue-colored band, a phenomenon previously reported in leaky-mode photonic lattices \cite{SGLee2020-1}. Moreover, with an increase in $h$, quasi-UUGRs and quasi-DUGRs also transition from the upper to lower band, as revealed in Figs.~\ref{fig3}(g) and \ref{fig3}(h). Along with these band transitions, we note a reduction in the maximum $\eta$ value. This leads to the reversion of the UUGR back to a quasi-UUGR with an $\eta$ value dropping below $80~\rm{dB}$. Concurrently, as illustrated in Fig.~\ref{fig3}(k), the $\eta$ values for quasi-DUGRs decrease, resulting in the formation of a DUGR with a pronounced downward radiation of $\eta = -81.6~\rm{dB}$ at $k_x/K = k_d=0.213862$ when $h=0.375~\Lambda$. The spatial electric field distributions, depicted in the inset of Fig.~\ref{fig3}(k), clearly demonstrate that the DUGR radiates only in the downward direction. The DUGR transitions back to a quasi-DUGR as the grating thickness $h$ increases beyond $h=0.375~\Lambda$. Furthermore, due to the in-plane $C_2$ symmetry, another DUGR was observed at $k_x/K =-0.213862$.

Figures~\ref{fig2} and \ref{fig3} illustrate that quasi-UGRs progressively transform into UGRs as the grating thickness $h$ is varied, while the grating width is kept constant at $w=0.4~\Lambda$. Through extensive FEM simulations, we have confirmed that this transition from quasi-UGRs to UGRs, occurring with variations in $h$, is quite common for different $w$ values. By closely tracking the evolution of quasi-UUGRs and quasi-DUGRs as $h$ varies, we have observed that genuine UUGRs and DUGRs emerge at different values of $h_u$ and $h_d$, respectively, for the different values of $w$. The relationship between $h_u$, $h_d$, and $w$, presented in the Supplementary Information, reveals that both UUGRs and DUGRs can be realized by gradually adjusting the grating thickness $h$ or the width $w$, while keeping the other parameter fixed. 

\begin{figure*}[]
\centering\includegraphics[width=16.5 cm]{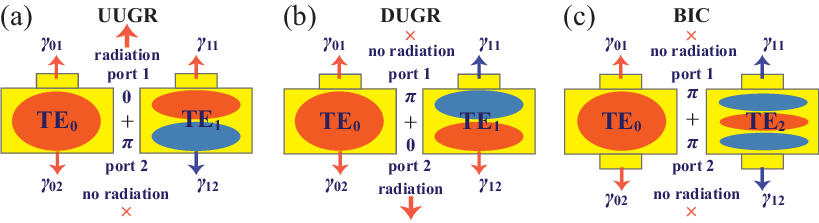}
\caption {\label{fig4} Conceptual illustrations depicting the formation mechanisms of (a) UUGRs, (b) DUGRs, and (c) BICs. UUGRs and DUGRs arise when radiating waves emitted from the even-like $\mathrm{TE}_{0}$ mode and odd-like $\mathrm{TE}_{1}$ mode undergo destructive interference in the downward and upward directions, respectively. Conversely, a coupled eigenstate manifests as a BIC with an infinite $Q$ factor when radiating waves from modes with identical transverse parities destructively interfere in both upward and downward directions simultaneously.}
\end{figure*}

\section{Discussion}
The formation of interband-coupling-induced UUGRs and DUGRs can be elucidated by examining the conceptual diagrams in Fig.~\ref{fig4}(a) for UUGRs and Fig.~\ref{fig4}(b) for DUGRs. As described by Eq.(\ref{eigenstates}), coupled eigenstates result from the combination of $\mathrm{TE}_0$ and $\mathrm{TE}_1$ modes. These modes respectively radiate in both upward (port 1) and downward (port 2) directions during the coupling process. Given that $\mathrm{TE}_0$ and $\mathrm{TE}_1$ modes possess even-like and odd-like transverse electric field distributions, the resulting radiating waves from these modes interfere constructively at one radiation port while destructively at the opposite port. As shown in Fig.~\ref{fig4}(a) (Fig.~\ref{fig4}(b)), coupled eigenstates become UUGRs (DUGRs) that radiate in the upward (downward) direction when the radiating waves from the even- and odd-like modes interfere destructively and cancel each other out in the downward (upward) direction. The process of forming UUGRs and DUGRs can be contrasted with the formation of BICs, as illustrated in Fig.~\ref{fig4}(c). When two waveguide modes having identical transverse parities, such as $\mathrm{TE}_0$ and $\mathrm{TE}_2$, are coupled, the resultant coupled eigenstates can form BICs if radiating waves from these modes destructively interfere at both radiation ports simultaneously.

Previous studies have demonstrated that topologically enabled UGRs arise when circularly-polarized $C$ points, which carry half-integer topological charges, coalesce in momentum space. To investigate the evolution of these $C$ points, it is crucial to calculate the polarization vectors of the far-field radiation in a 2D momentum space. Conventionally, this framework of topological charges was used to describe the formation of UGRs in isolated photonic bands. Extending this paradigm, Yin \emph{et al.} have recently demonstrated that the emergence of interband-coupling-induced UGRs can also be interpreted using this topological charge methodology \cite{XYin2023}. In our current study, we demonstrate that UUGRs and DUGRs induced by interband coupling between even-like and odd-like modes can be identified without direct reference to $C$ points. Instead, we track the evolution of quasi-UUGRs and quasi-DUGRs in 1D momentum space. Our approach, which utilizes the concept of quasi-UGRs, offers significant computational efficiencies. By conducting analyses in 1D momentum space, we are able to markedly reduce the computational demands. Nonetheless, we have confirmed that the topological charge framework in 2D momentum space is also valid for elucidating the formation of interband-coupling-induced UUGRs and DUGRs in ZCGs. A detailed exposition of the topological properties of UUGRs and DUGRs in ZCGs is provided in the Supplementary Information.

\begin{figure*}[]
\centering\includegraphics[width=16.0 cm]{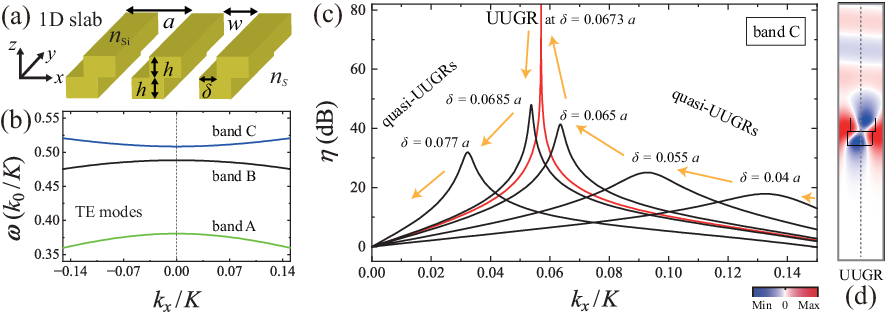}
\caption {\label{fig5} Quasi-UUGRs and genuine UUGRs in MDGs. (a) Schematic of an MDG configuration, composed of two misaligned  $\mathrm{Si}$ gratings embedded in a background medium with a refractive index of $n_s=1.46$. (b) Depiction of the lowest three photonic bands for $\mathrm{TE}$ modes, determined using FEM simulations. (c) The evolution of $\eta$ curves as a function of $k_x$ under variation of the grating misalignment $\delta$. (d) Visualization of the spatial $E_y$ field distribution for the UUGR at $k_x/K=0.05702$ confirms the unidirectional light emission in the upward direction.}
\end{figure*}

Next, we discuss the applicability of our 1D approach  in realizing topologically enabled UGRs in photonic lattices that lack both in-plane $C_2$ symmetry and up-down mirror symmetry. We consider a 1D misaligned dual grating (MDG), as depicted in Fig.~\ref{fig5}(a). This MDG consists of two misaligned $\mathrm{Si}$ gratings, each characterized by a lattice constant $a$, width $w=0.46~a$, and thickness $h=0.32~a$. The lateral misalignment, denoted as $\delta$, disrupts the up-down mirror symmetry while preserving inversion symmetry. Figure~\ref{fig5}(b) illustrates the three lowest $\mathrm{TE}$ photonic bands when $\delta =0$. Previous research within the topological charge framework has shown that various topological singularities, including UGRs, can be generated and manipulated in the $\mathrm{C}$ band by adjusting the misalignment $\delta$ \cite{YZeng2021}. As illustrated in Fig.~\ref{fig5}(c), the transition from quasi-UUGRs to UUGRs is evident by examining the evolution of radiation ratio $\eta$ curves as $\delta$ varies. A quasi-UUGR is generated and moves down along the dispersion curve $\mathrm{C}$ as $\delta$ increases from zero. Correspondingly, the $\eta$ values for quasi-UUGRs increases with $\delta$, leading to the emergence of a UUGR at $k_x/K=0.05702$ when $\delta=0.0673~a$. This UUGR reverts to a quasi-UUGR with a further increase in $\delta$ beyond $0.0673~a$. The spatial electric field distribution shown in Fig.~\ref{fig5}(d) confirms that this UUGR predominantly emits in the upward direction. Figure~\ref{fig5} reveals that quasi-UUGRs transition into a genuine UUGR for $k_x >0$. Additionally, due to the inversion symmetry inherent in the MDGs under study, a DUGR originating from quasi-DUGRs is observed at  $k_x/K=-0.05702$.

In summary, our study demonstrates that conventional 1D ZCGs can effectively function as unidirectional light emitters. In ZCGs, characterized by broken up-down mirror symmetry but retaining in-plane $C_2$ symmetry, the interband coupling between even-like and odd-like waveguide modes leads to the formation of quasi-UUGRs and quasi-DUGRs. These quasi-resonances are readily identifiable in 1D momentum space and exhibit notable robustness against variations in grating thickness. As the grating thickness increases, quasi-UUGRs and quasi-DUGRs evolve into genuine UUGRs and DUGRs, respectively. The significance of ZCG-based unidirectional light emitters lies in their ease of fabrication using existing nanofabrication technologies, making them highly relevant for practical applications. While previous research has emphasized tracking $C$ points in 2D momentum space to observe topologically enabled UGRs, our approach in 1D momentum space offers considerable reductions in computational time and resource demands. Furthermore, we have shown that this 1D methodology, centered around the evolution of quasi-UGRs, is applicable for characterizing UGRs in a variety of photonic lattice configurations, even those lacking both up-down mirror and in-plane $C_2$ symmetries. Our results offer substantial potential benefits for the advancement of various on-chip optoelectronic devices, including directional surface-emitting lasers, high efficiency grating couplers, and antennas for light detection.

\section{Materials and Methods}
In this study, all FEM simulations were conducted using the commercial software COMSOL Multiphysics. Given that the ZCGs under consideration are invariant in the $y$-direction, the simulations were carried out in a 2D $x$-$z$ plane using a computational unit cell with dimensions $\Lambda \times 10~\Lambda$, where $\Lambda$ represents the grating period. Bloch periodic boundary conditions were applied along the $x$-direction, while perfectly matched layers (PMLs) were utilized in the $z$-direction to absorb any outgoing radiation from the ZCGs. To ensure the reliability of the simulation results, a physics-controlled mesh with an extremely fine element size was employed. The eingenfrequency solver was implemented to calculate the eigenmodes, band structures, radiative $Q$ factors, and radiation ration $\eta$. In recent years, the application of FEM simulations has become increasingly prevalent in the exploration of topological physical phenomena. The detailed and accurate results obtained from FEM simulations provide critical insights and reliable guidelines for subsequent experimental implementations.  


\bigskip
\noindent\textbf{Acknowledgements}
This research was supported by the grant from the National Research Foundation of Korea, funded by the Ministry of Science and ICT (No. 2022R1A2C1011091) and the Ministry of Education (No. 2021R1I1A1A01060447).

\bigskip
\noindent\textbf{Author details}\\ 
(1) Department of Data Information and Physics, Kongju National University, Gongju, 32588, Republic of Korea. (2) Institute of Application and Fusion for Light, Kongju National University, Cheonan, 31080, Republic of Korea. (3) Advanced Photonics Research Institute, Gwangju Institute of Science and Technology, Gwangju 61005, Republic of Korea.

\bigskip
\noindent\textbf{Author contributions}\\
All author contributed substantially to this work.

\bigskip
\noindent\textbf{Conflict of Interest}\\
The authors declare no conflicts of interest.

\bigskip
\noindent\textbf{Data Availability Statement}\\
Data underlying the results in this paper may be obtained from the authors upon reasonable request.

\bigskip
\noindent\textbf{Supporting Information}\\
The online version contains supplementary materials available.


\end{document}